\documentclass[notitlepage,amssymb, nobibnotes, aps, prd,showpacs,showkeys]{revtex4-1}
\usepackage[english]{babel}
\usepackage{amsfonts}
\usepackage{amsthm}
\usepackage{graphicx}
\usepackage{braket}
\usepackage{bbold}
\usepackage{amssymb}
\usepackage{bbm}
\usepackage{subfigure}
\usepackage{hyperref}
\usepackage{amsmath}
\usepackage[usenames,dvipsnames,svgnames,table]{xcolor}
\usepackage{multirow}
\usepackage{axodraw}
\usepackage{bm}

\renewcommand{\epsilon}{\varepsilon}

\newcommand{\red}[1]{\textcolor{Red}{\bf #1}}
\newcommand{\mathred}[1]{{\color{Red}\bm{ #1}}}
\newcommand{\id}{\mathbb{1}}
\newcommand{\tr}{\mbox{Tr}}

\begin{document}

\title{Doubly Charmed Tetraquarks in $B_c$ and $\Xi_{bc}$ Decays}
\author{A. Esposito$^{*,\S}$,  M. Papinutto$^{*,\P}$, A. Pilloni$^{*,\P}$, A.D. Polosa$^{*,\P}$, N. Tantalo$^\dag$}
\affiliation{$^*$Dipartimento di Fisica, ``Sapienza'' Universit\`a di Roma, P.le A. Moro 2, I-00185 Roma, Italy\\ 
$^\S$Department of Physics, 538W 120th Street, Columbia University, New York, NY, 10027, USA\\
$^\P$INFN Sezione di Roma, Piazzale Aldo Moro 2, I-00185 Roma, Italy\\
 $^\dag$Dipartimento di Fisica and INFN, Universit\`a di Roma ``Tor Vergata'', Via della Ricerca Scientifica 1, I-00133 Roma, Italy}
 
\begin{abstract}
The phenomenology of the so-called $X$, $Y$ and $Z$ hadronic resonances
is hard to reconcile with standard charmonium or bottomonium interpretations. 
It has been suggested that some of these new hadrons can possibly be described as 
tightly bound tetraquark states and/or as loosely bound two-meson molecules. 
In the present paper we focus on the hypothetical existence of flavored, doubly charmed, tetraquarks. 
Such states might also carry double electric charge, and in this case, if discovered, 
they could univocally be interpreted in terms of compact tetraquarks.
Flavored tetraquarks are also amenable to lattice studies as their interpolating operators 
do not overlap with ordinary meson ones. 
We show that doubly charmed tetraquarks could  significantly be
produced at LHC from  $B_c$ or $\Xi_{bc}$ heavy baryons. 

\end{abstract}

\maketitle

\section{Introduction}
In the last decade, $B$-factories have discovered a considerable number of hadronic resonances
with hidden charm or beauty, the so-called $XYZ$ states~\cite{revs}, which do not fit standard quarkonium interpretations thereby being named as ``exotic''.  These observations opened a new field in hadron spectroscopy, which still lacks of a comprehensive theoretical framework. 
The new hadrons are interpreted  as compact tetraquark states~\cite{maiani}, or loosely bound meson molecules~\cite{mol} according to the most compelling phenomenological models proposed in the literature.  

Here we investigate the phenomenology of doubly charmed states~$\mathcal{T}$, whose experimental signatures would be very neat. 
In particular doubly charged ones could not be interpreted in terms of loosely bound molecules even if they occurred at threshold, because of the Coulomb barrier.

Doubly charmed particles have already been proposed in the literature~\cite{Moinester:1995tt,DelFabbro:2004ta,Valcarce:2010zs,Hyodo:2012pm,Bicudo:2012qt}, focusing especially on states with masses lying below their relative open charm threshold. In this paper we make the opposite hypothesis that~$\mathcal{T}$ states might indeed occur above threshold and be produced in $B_c$ and bottom-charmed baryon decays at LHC. 

From a theoretical point of view, the flavor quantum numbers of the $\mathcal{T}$ give the possibility to clearly distinguish these tetraquarks from conventional quarkonia and, at the same time, allow lattice simulations to make predictions, even if these particles are not QCD stable but their mass is sufficiently close to the open charm threshold.

This is one of the main motivations of our work as it will be discussed in the next section. At this stage of our program, beyond setting the frame for future lattice investigations, we focus on the phenomenology of $\mathcal{T}$ particles providing a first ``hunter guide'' for their searches at the LHC.  In Section~\ref{sec:tetraq} we discuss decay modes and production mechanisms of $\mathcal{T}$ particles in a constituent diquark-antidiquark model. We estimate decay widths and production branching fractions  from $B_c$ mesons. A Section on production from $\Xi_{bc}$ baryons is also included. More details on the estimates here obtained can be found in the Appendices together with a discussion on our ongoing lattice program. 

\section{Theoretical Motivations}
Hadron masses are
obtained by studying the euclidean-time behaviour of correlators of
suitably chosen interpolating operators
\begin{equation}
\langle O^\dag_H(t) O_H(0)\rangle =
\sum_n \vert \bra{H_n} \hat O_H  \ket{0}\vert^2\ e^{-E_n t} 
\label{eq:opemap}
\end{equation}
where $\hat O_H$ represents the interpolating operator $O_H$ in the Hamiltonian
formalism. In absence of electroweak interactions, hadrons can be identified with the
eigenstates $\ket{H_n}$ of the QCD Hamiltonian, and the hadron spectrum with the corresponding eigenvalues $E_n$. 
Interpolating operators $\hat O_H$  with different quantum numbers give
access to different sectors of the QCD spectrum. 

On the other hand, operators with the \emph{same} quantum numbers interpolate the same
states. For example, both the operators
\begin{align}
\hat O_p &= \varepsilon^{ijk} u^i  u^j  d^k & 
\hat O_p^{\;\prime} &= \varepsilon^{ijk}  u^i  u^j  d^k\ \bar s^h s^h 
\end{align}
(for the sake of simplicity we display only the color and flavor indices and not the Dirac 
structure) can create from the vacuum a baryon with definite quantum numbers 
but also all his excited states and many-particle states.
The difference between the two operators consists in the strength of
the coupling to the ground state ({\it i.e.}, with suitable Dirac structure, the proton) 
and to the excited states. 

Consider now an hypothetic state $\ket{H^0_{\bar c c}}$ with  flavor
content $[\bar c c] [\bar u u]$. The molecular hypothesis~\cite{mol} and the diquark-antidiquark model~\cite{maiani} would suggest to use  
\begin{align}
\hat O_{mol} &= \bar c^i u^i\ \bar u^j c^j &
&\text{or} & 
\hat O_{diq} &= \varepsilon^{ijk} \bar c^j \bar u^k\ \varepsilon^{ilm} c^l u^m 
\end{align}
as interpolating operators of $\ket{H^0_{\bar c c}}$. 
However 
$\hat O_{mol}$ and $\hat O_{diq}$ have in fact the same quantum numbers. Furthermore, the state
$\ket{H^0_{\bar c c}}$ can also be created by using flavor singlet meson operators such as
\begin{equation}
\hat O_{mes}= \left\{\bar c^i c^i,\ \bar u^i u^i,\ \bar s^i s^i,\ \cdots \right\}
\end{equation}

All these objects mix under renormalization. In practice this means that, although there may
be a certain combination of operators $\hat O^\star$ that ``maximizes'' the overlap 
with the state $\ket{H^0_{\bar c c}}$, the value of the matrix element $\bra{H^0_{\bar c c}} \hat O^\star \ket{0}$
would unavoidably depend upon the scale and the scheme used in the renormalization procedure.
The concept of valence quarks can be extended to quantum field theory by defining the ``number of quarks'' 
as the eigenvalues of the operator
\begin{align}
\hat N_{val} & =\sum_f \vert \hat Q_f \vert &
\hat Q_{f} &= \int d^3x\ \bar \psi_f^i \gamma^0   \psi_f^i &
f &=\left\{u,d,s,c,b,t\right\}
\label{eq:nquarkdef}
\end{align}
According to this definition, mesons can have either $N_{val}=2$ or 
$N_{val}=0$. The state $\ket{H^0_{\bar c c}}$ would have $N_{val}=0$ and, consequently, would be hardly distinguishable from a conventional flavor singlet quarkonium state.

In this paper we consider states with $N_{val}=4$, {\it i.e.}  
flavor quantum numbers which necessarily require four valence quarks and cannot be created by a meson interpolating operator with further addition of flavor singlet quark-antiquark pairs. For this reason in the following we refer to $N_{val}=4$ particles as to ``pure tetraquarks''.

If a pure tetraquark state is discovered and it has a mass below the lightest two meson threshold with the 
appropriate flavor quantum numbers, it would be stable against  (flavor conserving) strong and electromagnetic interactions. This observation has already been made in the literature~\cite{Moinester:1995tt,DelFabbro:2004ta,Valcarce:2010zs,Hyodo:2012pm,Bicudo:2012qt}. Stable tetraquarks have been conjectured because they would correspond to the smallest eigenvalue of the QCD Hamiltonian with the prescribed quantum numbers and, for this reason, it would be 
a relatively easy task to search for them on the lattice by using Eq.~\eqref{eq:opemap}. See also the large-$N$ analyses in Refs.~\cite{weinberg,peris}.

However, pure tetraquarks do not necessarily need to be stable, which is the case we are most interested in. Despite the fact that it is not possible to deal with generic hadronic resonances on the lattice, the peculiar flavor structure of pure tetraquarks makes a lattice calculation affordable by using L\"uscher's method~\cite{Luscher:1986pf} as discussed in Appendix~\ref{sec:applatt}.

In the coming sections we perform a phenomenological study on possible quantum numbers, decay modes and production rates of doubly charmed pure tetraquark states. We shall rely on the constituent diquark-antidiquark model. 

\section{The tetraquark model}
\label{sec:tetraq}
\subsection{Spectroscopy}
Within the constituent diquark-antidiquark model~\cite{exotica}, first used to describe heavy-light tetraquarks in~\cite{maiani},  the existence of the doubly charmed  pure tetraquark states
discussed in the previous section
\begin{equation}
 \mathcal{T} = [cc][\bar q_1 \bar q_2] \text{, with } q_1,q_2=u,d,s
\end{equation}
is predicted. 
We assume that the two quarks (antiquarks) combine in the $\bm{\bar{3}}_c$ ($\bm{3}_c$) configuration, which is attractive in one-gluon-exchange approximation. Moreover, the total wave function of such a diquark (antidiquark) has to be antisymmetric because of Fermi statistics. As for the $[cc]$ diquark, the flavor is forced to be symmetric, and so we can only have
\begin{equation}
 [cc] = \left| \bm{\bar{3}}_c(A), \,J^{P}\!=\!1^+(S)\right\rangle
\end{equation}
where by $(S)$ and $(A)$ we indicate the symmetry/antisymmetry of a configuration.

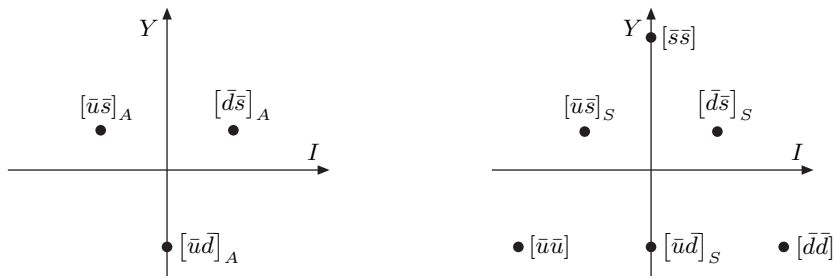
\begin{figure}[h]
\centering
\begin{picture}(180,100)(-30,0)
\SetWidth{0.5}
\Line(60,0)(60,100)
\Line(0,40)(120,40)
\ArrowLine(60,98)(60,100)
\ArrowLine(118,40)(120,40)
\Vertex(85,55){2}
\Vertex(35,55){2}
\Vertex(60,11){2}
\Text(77,59)[bl]{$\left[\bar d \bar s\right]_A$}
\Text(27,59)[bl]{$\left[\bar u \bar s\right]_A$}
\Text(64,5)[bl]{$\left[\bar u \bar d\right]_A$}
\Text(118,44)[br]{$I$}
\Text(57,97)[tr]{$Y$}
\end{picture}
\begin{picture}(180,100)(-30,0)
\SetWidth{0.5}
\Line(60,0)(60,100)
\Line(0,40)(120,40)
\ArrowLine(60,98)(60,100)
\ArrowLine(118,40)(120,40)
\Vertex(85,54.5){2}
\Vertex(35,54.5){2}
\Vertex(60,90){2}
\Vertex(10,11){2}
\Vertex(60,11){2}
\Vertex(110,11){2}
\Text(77,59)[bl]{$\left[\bar d \bar s\right]_S$}
\Text(27,59)[bl]{$\left[\bar u \bar s\right]_S$}
\Text(64,5)[bl]{$\left[\bar u \bar d\right]_S$}
\Text(14,11)[cl]{$[\bar u \bar u]$}
\Text(115,11)[cl]{$[\bar d \bar d]$}
\Text(64,90)[cl]{$[\bar s \bar s]$}
\Text(118,44)[br]{$I$}
\Text(57,97)[tr]{$Y$}
\end{picture}
\caption{Quark content of the good (left) and bad (right) light antidiquark. $A$ and $S$ stand for the antisymmetric and symmetric flavor combinations. The good configuration contains one isospin doublet and one isospin singlet, while the bad configuration also has an isospin triplet. Note that the $\left[ cc\right] \left[ \bar u \bar u\right]$ corresponds to an electrically neutral particle with $I_3 \neq 0$. }
\label{fig:multiplet}
\label{fig:diquarks}
\end{figure}

As for the light antidiquark, we have two choices ($f$ is referred to flavor space)
\begin{subequations}
\begin{align}
 [\bar q_1 \bar q_2]_G &= \left| \bm{3}_c(A), \,\bm{3}_f(A) ,\,J^{P}\!=\!0^+(A)\right\rangle\label{eq:good}\\
 [\bar q_1 \bar q_2]_B &= \left| \bm{3}_c(A), \,\bm{\bar{6}}_f(S),\, J^{P}\!=\!1^+(S)\right\rangle\label{eq:bad}
\end{align}
\end{subequations}
Here the $\bm 3$ representations are treated as odd, $(A)$, being the antisymmetric part of the tensorial product \mbox{$\bm{\bar{3}}\otimes\bm{\bar{3}}$}.
According to the phenomenological color-spin Hamiltionian, the $G$ (``good'') scalar state in Eq.~\eqref{eq:good} is expected to be lighter (hence most likely produced) than the $B$ (``bad'') vectorial state in Eq.~\eqref{eq:bad}. In Fig.~\ref{fig:multiplet} we show the light diquark flavor multiplets.

The $ \mathcal{T}$ particles can be formed as neutral, charged and doubly charged states. As we remarked in the introduction, the doubly charged configurations are particularly interesting because, even if one of them were experimentally discovered close to any open charm threshold, it should unavoidably be interpreted as a compact tetraquark. Indeed, in a loosely bound molecule, an extended object with respects to the range of strong interactions, Coulomb repulsion would induce a fall apart decay on very short timescales. 

We can now combine the $1^+$ $[cc]$ diquark with both good and bad light antidiquarks, to obtain good and bad $\mathcal{T}$ states, as in Tab.~\ref{tab:states}. As for quantum numbers, a good $\mathcal{T}$ can be only $1^+$, while a bad $\mathcal{T}$ can have $J=0,1,2$. Anyway, we expect the scalar configuration to be lighter and more likely to be formed. 
\begin{table}[h]
\begin{tabular}{c|c}
\hline\hline
\multicolumn{2}{c}{$\mathcal{T}$ states}\\
\hline
``Good'', $\red{1^+}$ & ``Bad'', $\red{0^+},1^+,2^+$\\
\hline
$\mathcal{T}^+\; ([cc]{[\bar u\bar d\,]_A})$ & $\mathcal{T}^0\; ([cc][\bar u\bar u])$\\
$\mathcal{T}^+_s\; ([cc][\bar u \bar s]_A)$ & $\mathcal{T}^{++}\; ([cc][\bar d\bar d])$\\
$\mathcal{T}^{++}_s\; ([cc][\bar d\bar s]_A)$ & $\mathcal{T}^{++}_{ss}\; ([cc][\bar s\bar s])$\\
& $\mathcal{T}^+\; ([cc]{[\bar u\bar d\,]_S})$\\
& $\mathcal{T}^+_s\; ([cc][\bar u \bar s]_S)$ \\
& $\mathcal{T}^{++}_s\; ([cc][\bar d\bar s]_S)$\\
\hline\hline
\end{tabular}
\caption{Expected $\mathcal{T}$ states. $A$ and $S$ stand for the antisymmetric and symmetric flavor combinations. The configurations in \red{red} are those more likely formed. }
\label{tab:states}
\end{table}

\subsection{Strong Decays}
Let us focus on the $\mathcal{T}_s^{++}$ (see Tab.~\ref{tab:states}), being the extension to the other states straightforward. The allowed decay channels depend crucially on whether the $\mathcal{T}$ state is above the open charm threshold. If not, strong decays are forbidden. The weak decay channels would present a too complicated pattern that would make the experimental analysis too challenging. Therefore, in the following we limit our discussion to above-threshold states. As for the decay products, we take into account only the lightest $0^-$ and $1^-$ open charm mesons. Hence, we consider only $S$-wave decays, because $P$-wave decays are parity forbidden. The decay channels are reported in Tab.~\ref{tab:decays}.
\begin{table}[h]
\begin{tabular}{c|c|c|c}
\hline\hline
\multicolumn{4}{c}{$\mathcal{T}_s^{++}$ decays}\\
\hline
$\red{0^+}$ \red{bad}& $\red{1^+}$ \red{good} & $1^+$ bad& $2^+$ bad\\
\hline
$D_s^+D^+$ & $D_s^{*+}D^+$ & $D_s^{*+}D^+$ & $D_s^{*+}D^{*+}$\\
$D_s^{*+}D^{*+}$ & $D_s^+D^{*+}$ & $D_s^+D^{*+}$ & \\
& $D_s^{*+}D^{*+}$ & & \\
\hline\hline
\end{tabular}
\caption{Possible $\mathcal{T}_s^{++}$ decays channels. The configurations in \red{red} are those more likely produced. }
\label{tab:decays}
\end{table}

We see that a $1^+$ $\mathcal{T}_s^{++}$ cannot decay into two pseudoscalar mesons. If one takes into account only good configurations, then we have a selection rule. A similar situation arises for every $\mathcal{T}$ made by different light quarks.

On the other hand, when the light quarks are identical (as in $\mathcal{T}^0$, $\mathcal{T}^{++}, \mathcal{T}_{ss}^{++}$), there is no good configuration. If we limit to the most likely $0^+$ bad state, it cannot decay into a $PV$ configuration. Finally, we remark that the bad $1^+$ state cannot decay into two vectors because of heavy quark spin conservation.

\begin{figure}[b]
\centering
 \includegraphics[width=.45\textwidth]{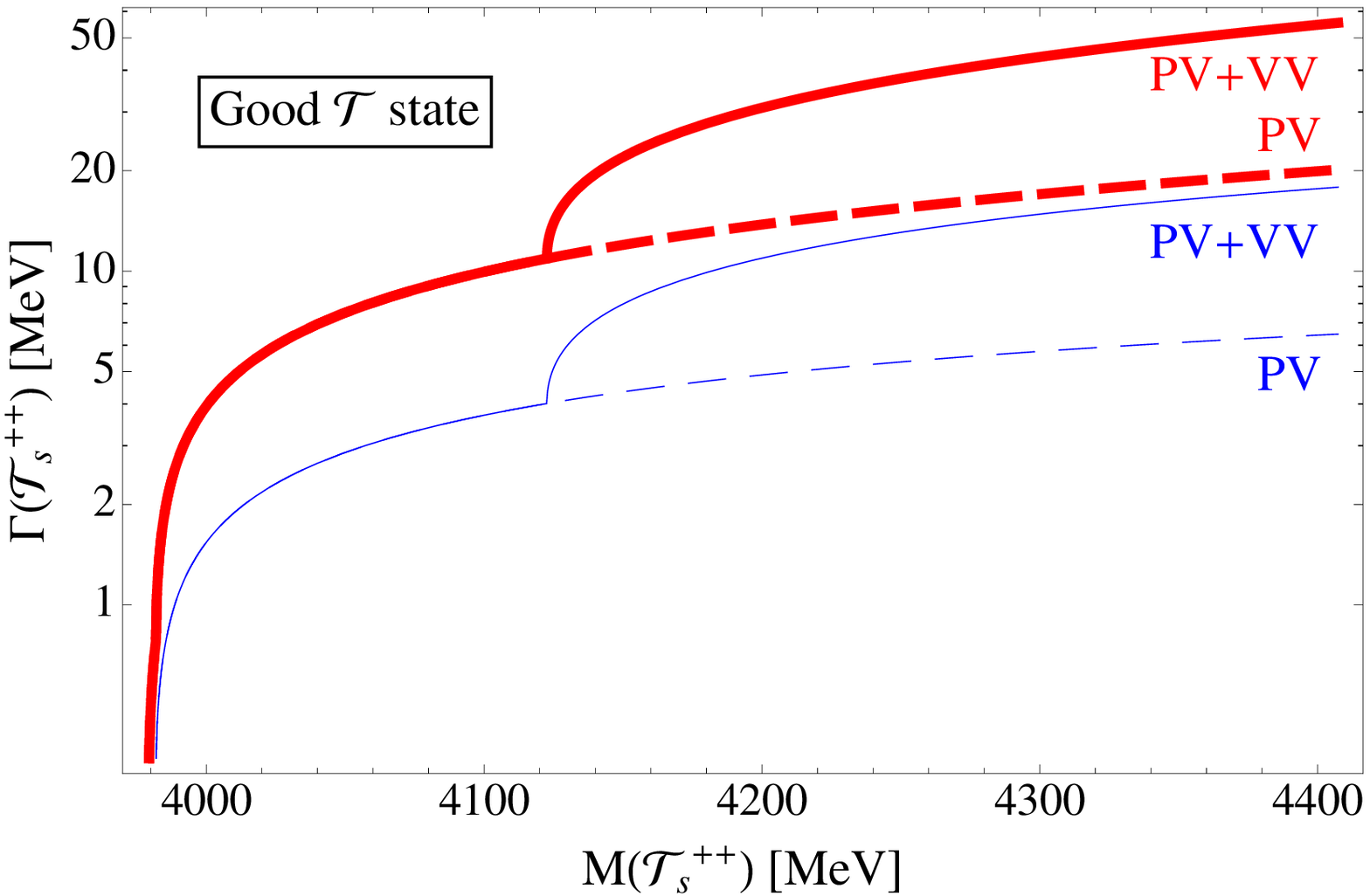}
 \hspace{1em}
 \includegraphics[width=.45\textwidth]{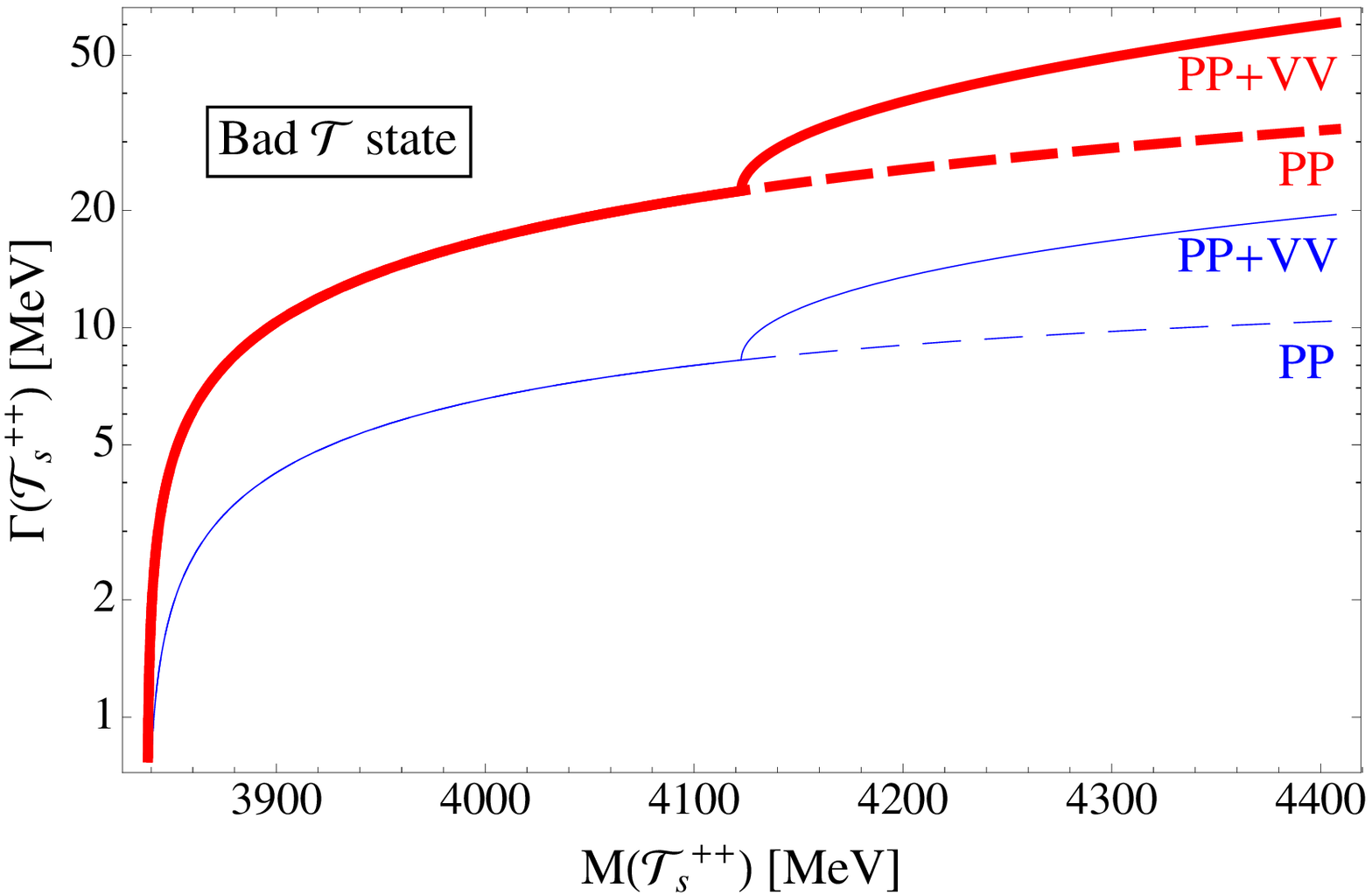}
 \caption{Width of good $1^+$ (left) and bad $0^+$ (right) $\mathcal{T}$ as a function of the mass for $g_\mathcal{T}=M_\mathcal{T}$ (thick) and $g_\mathcal{T}=g_{XDD^*}$~(thin). With $P$ and $V$ we indicate the final states $D_{(s)}$ and $D^*_{(s)}$ respectively. }
 \label{fig:widths}
\end{figure}

In order to estimate the width, we parametrize the decay matrix elements as
\begin{subequations}
\begin{align}
\left\langle \mathcal{T}_s^{(G)++}\left(P,\epsilon\right)\,\Big|\, D_s^{+}\left(p\right)\,D^{*+}\left(q,\lambda\right)\right\rangle &= \frac{g_\mathcal{T}}{2}\,\epsilon \cdot \lambda\\
\left\langle \mathcal{T}_s^{(G)++}\left(P,\epsilon\right)\,\Big|\, D_s^{*+}\left(p,\eta\right)\,D^{+}\left(q\right)\right\rangle &= \frac{g_\mathcal{T}}{2}\,\epsilon \cdot \eta\\
\left\langle \mathcal{T}_s^{(G)++}\left(P,\epsilon\right)\,\Big|\, D_s^{*+}\left(p,\eta\right)\,D^{*+}\left(q,\lambda\right)\right\rangle &=\frac{g_\mathcal{T}}{\sqrt{2}}\,\epsilon_{\mu\nu\rho\sigma} P^\mu \epsilon^\nu \eta^\rho \lambda^\sigma\\
\left\langle \mathcal{T}_s^{(B)++}\left(P\right)\,\Big|\, D_s^{+}\left(p\right)\,D^{+}\left(q\right)\right\rangle &= \frac{\sqrt{3}}{2}g_\mathcal{T}\\
\left\langle \mathcal{T}_s^{(B)++}\left(P\right)\,\Big|\, D_s^{*+}\left(p,\eta\right)\,D^{*+}\left(q,\lambda\right)\right\rangle &=\frac{g_\mathcal{T}}{2}\,\eta\cdot\lambda
\end{align}
\end{subequations}
where $g_\mathcal{T}$ is an effective strong coupling with the dimension of a mass, and the numerical factors take into account the Clebsch-Gordan coefficients for the Fierz transformation needed to rearrange the diquark color structure into open charm mesons. In order to compute the total width we must assign a value to the unknown $g_\mathcal{T}$. In particular, we could assume $g_\mathcal{T}\simeq M_\mathcal{T}$ by dimensional analysis, or we could use the same coupling as the $X(3872)\to D D^*$ one, which might also be considered a tetraquark decay into open charm mesons, that is $g_\mathcal{T}\simeq g_{XDD^*}=2.5$ GeV~\cite{Faccini:2013lda}. In Fig.~\ref{fig:widths} we plot the total widths for both the choices of the coupling, and for both good ($1^+$) and bad ($0^+$) configurations, as a function of the mass. The resulting total widths are of the same order of magnitude. In particular, they are narrow enough to be experimentally measured. 

\subsection{\texorpdfstring{Production from the $B_c^+$}{Production from the Bc}}

\begin{figure}[t]
\centering
\begin{picture}(150,80)(0,0)
\SetWidth{0.5}
\ArrowLine(0,0)(120,0)
\Text(-7,0)[cl]{$c$}
\Text(127,0)[cr]{$c$}
\ArrowLine(120,80)(0,80)
\Text(-7,80)[cl]{$\bar b$}
\Text(127,80)[cr]{$\bar c$}

\Photon(40,80)(80,55){1}{6}
\Gluon(40,0)(80,25){2}{7}

\ArrowLine(80,55)(120,65)
\ArrowLine(120,45)(80,55)
\Text(127,66)[cr]{$c$}
\Text(127,45)[cr]{$\bar s$}

\ArrowLine(80,25)(120,35)
\ArrowLine(120,15)(80,25)
\Text(123,30)[bl]{$u,d,s$}
\Text(123,22)[tl]{$\bar u, \bar d, \bar s$}
\Text(38,73)[cc]{$\lambda^2$}
\end{picture}
\begin{picture}(150,80)(0,0)
\SetWidth{0.5}
\ArrowLine(0,0)(120,0)
\Text(-7,0)[cl]{$c$}
\Text(127,0)[cr]{$c$}
\ArrowLine(120,80)(0,80)
\Text(-7,80)[cl]{$\bar b$}
\Text(127,80)[cr]{$\bar c$}

\Photon(40,80)(80,55){1}{6}
\Gluon(40,0)(80,25){2}{7}

\ArrowLine(80,55)(120,65)
\ArrowLine(120,45)(80,55)
\Text(127,66)[cr]{$c$}
\Text(127,45)[cr]{$\bar d$}

\ArrowLine(80,25)(120,35)
\ArrowLine(120,15)(80,25)
\Text(123,30)[bl]{$u,d,s$}
\Text(123,22)[tl]{$\bar u, \bar d, \bar s$}
\Text(38,73)[cc]{$\lambda^2$}
\Text(72,50)[cc]{$\lambda$}

\end{picture}
\begin{picture}(140,80)(0,0)
\SetWidth{0.5}
\ArrowLine(0,0)(120,0)
\Text(-7,0)[cl]{$c$}
\Text(127,0)[cr]{$c$}
\ArrowLine(120,80)(0,80)
\Text(-7,80)[cl]{$\bar b$}
\Text(127,80)[cr]{$\bar u$}

\Photon(40,80)(80,55){1}{6}
\Gluon(40,0)(80,25){2}{7}

\ArrowLine(80,55)(120,65)
\ArrowLine(120,45)(80,55)
\Text(127,66)[cr]{$c$}
\Text(127,45)[cr]{$\bar s$}

\ArrowLine(80,25)(120,35)
\ArrowLine(120,15)(80,25)
\Text(123,30)[bl]{$u,d,s$}
\Text(123,22)[tl]{$\bar u, \bar d, \bar s$}
\Text(38,73)[cc]{$\lambda^3$}

\end{picture}
\caption{Double- and triple-Cabibbo suppressed Feynman diagrams for the production of the $\mathcal{T}$ particles from $B_c^+$.}\label{diagrammi_mesoni}
\end{figure}
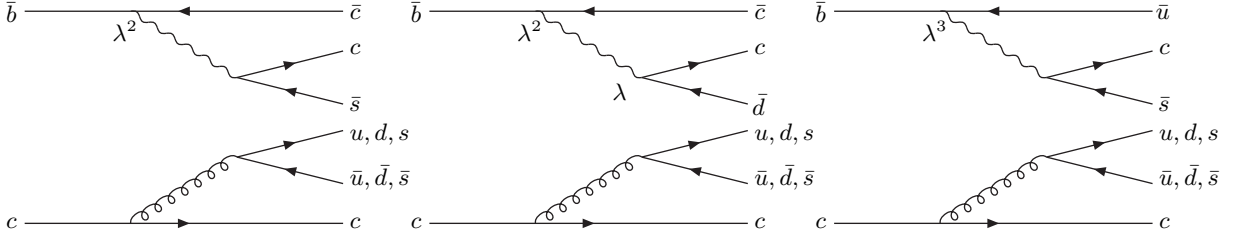

We expect $\mathcal{T}$ particles to be produced in $B_c$ meson decays. In this situation we have three different kinds of process, of order $\mathcal{O}(\lambda^2)$ and $\mathcal{O}(\lambda^3)$ ($\lambda$ is the Cabibbo angle), depending on the bottom quark decay channel: $\bar b\to \bar c c\bar s$, $\bar b\to \bar c c \bar d$ and $\bar b\to \bar u c \bar s$. The Feynman diagrams for such processes are shown in Fig.~\ref{diagrammi_mesoni}. The possible final products of the $B_c^+$ decays involving $\mathcal{T}$ states are presented in Tab.~\ref{table_Bc}.

We want to provide an estimate of the branching ratios for the production of the $\mathcal{T}$ particles, \emph{e.g.} $B_c^+\to \mathcal{T}_s^{++}D^{(*)-}$, avoiding specific models. 
We found that heavy meson decays in two baryons are particularly suitable to extract an effective strong coupling which we might expect to determine also the process  we are interested in -- both indeed contain six  quarks confined in a  two-hadron final state. 
In particular, we choose
\begin{equation}
\mathcal{BR}\left(B^0\to \bar{\Lambda}_c^-\, p\right) = \left(2.0 \pm 0.4\right)\times 10^{-5} \,\text{ \cite{lambdac}} ,\qquad \mathcal{BR}\left(B^+\to \bar{\Sigma}_c^0\, p\right) = \left(3.7 \pm 1.5\right)\times 10^{-5} \,\text{ \cite{sigmac}}
\end{equation}

\begin{figure}[t]
\centering
\begin{picture}(180,80)(0,-2)
\SetWidth{0.5}
\ArrowLine(0,0)(120,0)
\Text(-8,0)[cl]{$c_i$}
\Text(122,0)[cl]{$c^\prime_j$}
\ArrowLine(120,80)(0,80)
\Text(-8,80)[cl]{$\bar b_i$}
\Text(130,80)[cr]{$\bar c_i$}

\Photon(40,80)(80,55){1}{6}
\Gluon(40,0)(80,25){2}{7}

\ArrowLine(80,55)(120,65)
\ArrowLine(120,45)(80,55)
\Text(123,60)[bl]{$c_k$}
\Text(123,51)[tl]{$\bar s_l$}

\ArrowLine(80,25)(120,35)
\ArrowLine(120,15)(80,25)
\Text(123,30)[bl]{$d_m$}
\Text(123,22)[tl]{$\bar d_n$}
\end{picture}
\begin{picture}(160,80)(0,-2)
\SetWidth{0.5}
\ArrowLine(0,0)(120,0)
\Text(-2,0)[cr]{$d_i\,(u_i)$}
\Text(122,0)[cl]{$d_j\,(u_j)$}
\ArrowLine(120,80)(0,80)
\Text(-8,80)[cl]{$\bar b_i$}
\Text(130,80)[cr]{$\bar c_i$}

\Photon(40,80)(80,55){1}{6}
\Gluon(40,0)(80,25){2}{7}

\ArrowLine(80,55)(120,65)
\ArrowLine(120,45)(80,55)
\Text(123,60)[bl]{$u_k$}
\Text(123,53)[tl]{$\bar d_l$}

\ArrowLine(80,25)(120,35)
\ArrowLine(120,15)(80,25)
\Text(123,29)[bl]{$u^\prime_m\,(d_m)$}
\Text(123,20)[tl]{$\bar u_n\,(\bar{d}^{\,\prime}_n)$}
\end{picture}
\caption{Feynman diagram for the processes  $B_c^+\to \mathcal{T}_s^{++}D^{(*)-}$ (left) and $B^0\to \bar{\Lambda}^-_c p \left(B^+\to \bar{\Sigma}^0_c p\right) $ (right). The indices represent the color of the quarks.}\label{fig:produzione_barioni}
\end{figure}
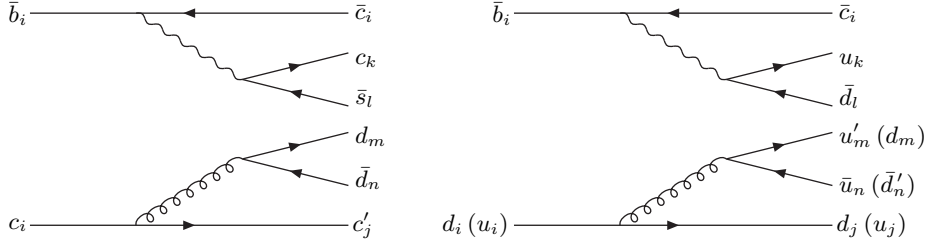

In Fig.~\ref{fig:produzione_barioni} the Feynman diagrams for such decays are shown. We assume the physics behind the three processes to be the same. In particular, we can parametrize these partial widths in terms of phase space, kinematical and color structure, and of an effective strong coupling which we assume to be the same in all the considered processes.

\begin{table}[b]
\centering
\begin{tabular}{ccc}
\hline\hline
& Bottom quark decays & \\
\hline
$\bar b\to \bar c c\bar s$\hspace{1em} $(\mathcal{O}(\lambda^2))$ &  $\bar b\to \bar c c \bar d$ \hspace{1em} $(\mathcal{O}(\lambda^3))$& $\bar b\to \bar u c \bar s$\hspace{1em} $(\mathcal{O}(\lambda^3))$ \\
\hline
$\bar{D}^{(*)0} \mathred{\mathcal{T}^+}\to \bar{D}^{(*)0} \mathred{D_s^{+} D^{0}}$ & $\bar{D}^{(*)0}\mathred{\mathcal{T}^+}\to \bar{D}^{(*)0}\mathred{D^0D^+}$ & $K^{(*)+}\mathcal{T}^0\to K^{(*)+}D^0D^0$ \\
$D^{(*)-}\mathred{\mathcal{T}_s^{++}}\to D^{(*)-}\mathred{D^+D_s^+}$ & $D^{(*)-}\mathcal{T}^{++}\to D^{(*)-}D^+D^+$ & $K^{(*)0}\mathred{\mathcal{T}^+}\to K^{(*)0}\mathred{D^0D^+}$\\
$D_s^{(*)-}\mathcal{T}_{ss}^{++}\to D_s^{(*)-}D_s^+D_s^+$ & $D_s^{(*)-}\mathred{\mathcal{T}^{++}_s}\to D_s^{(*)-}\mathred{D^+D_s^+}$ & $\phi\mathred{\mathcal{T}_s^+}\to\phi \mathred{D^0D_s^+}$ \\
& & $\pi^0(\rho^0)\mathred{\mathcal{T}_s^+}\to\pi^0(\rho^0)\mathred{D^0D_s^+}$\\
& & $K^{(*)-}\mathcal{T}^{++}_{ss}\to  K^{(*)-}D_s^+D_s^+$ \\
& & $\pi^-(\rho^-)\mathred{\mathcal{T}_s^{++}}\to\pi^-(\rho^-)\mathred{D_s^+D^+}$ \\
\hline\hline
\end{tabular}
\caption{Production channels of the $\mathcal{T}$ particles from $B_c^+$ decay. With $\mathcal{O}(\lambda^n)$ we indicate the power of the Cabibbo suppression of the considered process. It is understood that each meson can also be found in its excited state, depending on the $J^P$ quantum numbers of the $\mathcal{T}$ particle considered -- see Tab.~\ref{tab:decays}. We have marked in \red{red} good tetraquarks, and their decay products. In this case at least one charmed meson must be vectorial.}
\label{table_Bc}
\end{table}

As an example, let us focus on the process $B_c^+\to \mathcal{T}_s^{++}D^{(*)-}$, whose Feynman diagram is shown in Fig.~\ref{fig:produzione_barioni}, being the generalization to other $\mathcal{T}$ states straightforward. Using $SU(3)_c$ algebra it is possible to calculate the color factor of such process.
Factorizing the color structure, the width reads
\begin{equation}\label{eq:width_tantalo}
\Gamma(B_c^+\to\mathcal{T}_s^{++}D^{(*)-})=\frac{p^*(M_{B_c^+},M_{\mathcal{T}},M_{D^{(*)-}})}{8\pi M^2_{B_c^+}}c_1^2{\left|\mathcal{A}_{B_c^+}\right|}^2
\end{equation}
where $p^*$ is the decay 3-momentum and $c_1 = 4/81$ is the color factor evaluated in  Appendix~\ref{app:calcoli}.
As for the $B^0\to\bar{\Lambda}_c^-p$ process, instead, the width is
\begin{equation}\label{eq:B0}
\Gamma(B\to\bar{\Lambda}\,p)=\frac{p^*(M_{B},M_{\Lambda},M_{p})}{8\pi M_{B}^2}c_2^2{\left|\mathcal{A}_{B}\right|}^2
\end{equation}
with $c_2=2/81$, and $B=B^0\,\left(B^+\right)$, $\bar \Lambda=\bar \Lambda_c^-\,\left(\bar \Sigma_c^0\right)$.

We now need to express the various amplitudes in terms of the effective strong coupling and of the kinematical structure. Let us first analyze the baryonic decays. The effective lagrangian for such a reaction can be chosen to be analogous to the heavy meson chiral lagrangian~\cite{casalbuoni}
\begin{equation}\label{eq:lagrangiana}
\mathcal{L}_{\text{eff}}=\frac{g_B}{2M_B^2}\,\partial_\mu B\, \bar{p}\,\gamma^\mu\left(1-\frac{g_A}{g_V}\gamma_5\right)\Lambda
\end{equation}
where $g_B$ is a strong effective coupling with the dimension of a mass, for $g_A/g_V$ we choose the same as in the case of the neutron $\beta$-decay, $g_A/g_V \simeq 1.27$, and negative powers of $M_B$ have been added for dimensional reasons. In our situation the field $\Lambda$ represents both the $\bar{\Lambda}_c^-$ and the $\bar{\Sigma}_c^0$, the kinematical structure being the same for the two decays. Analogously, $B$ represents both the $B^0$ and the $B^+$.
From Eq.~\eqref{eq:lagrangiana} one finds the square amplitudes for the $B^0 \to \bar \Lambda_c^-\,p$ and $B^+\to \bar{\Sigma}_c^0\,p$ decays to be
\begin{subequations}\label{eq:matrix}
\begin{align}
{\left|\mathcal{A}_{B^0}\right|}^2&=\frac{g_{B^0}^2}{M_{B^0}^4}\left[\frac{M_{B^0}^2-M_{\Lambda_c^-}^2-M_p^2}{2}\left(\Delta M^2+\left(\frac{g_A}{g_V}\right)^2M^2\right)+\left(\left(\frac{g_A}{g_V}\right)^2M^2-\Delta M^2\right)M_p M_{\Lambda_c^-}\right] \\
{\left|\mathcal{A}_{B^+}\right|}^2&=\frac{g_{B^+}^2}{M_{B^+}^4}\left[\frac{M_{B^+}^2-M_{\Sigma_c^0}^2-M_p^2}{2}\left(\Delta M^2+\left(\frac{g_A}{g_V}\right)^2M^2\right)+\left(\left(\frac{g_A}{g_V}\right)^2M^2-\Delta M^2\right)M_p M_{\Sigma_c^0}\right]
\end{align}
\end{subequations}
where we have set $\Delta M=M_p-M_{\Lambda_c^-(\Sigma_c^0)}$ and $M=M_p+M_{\Lambda_c^-(\Sigma_c^0)}$.

\begin{figure}[b]
\centering
\includegraphics[scale=0.5]{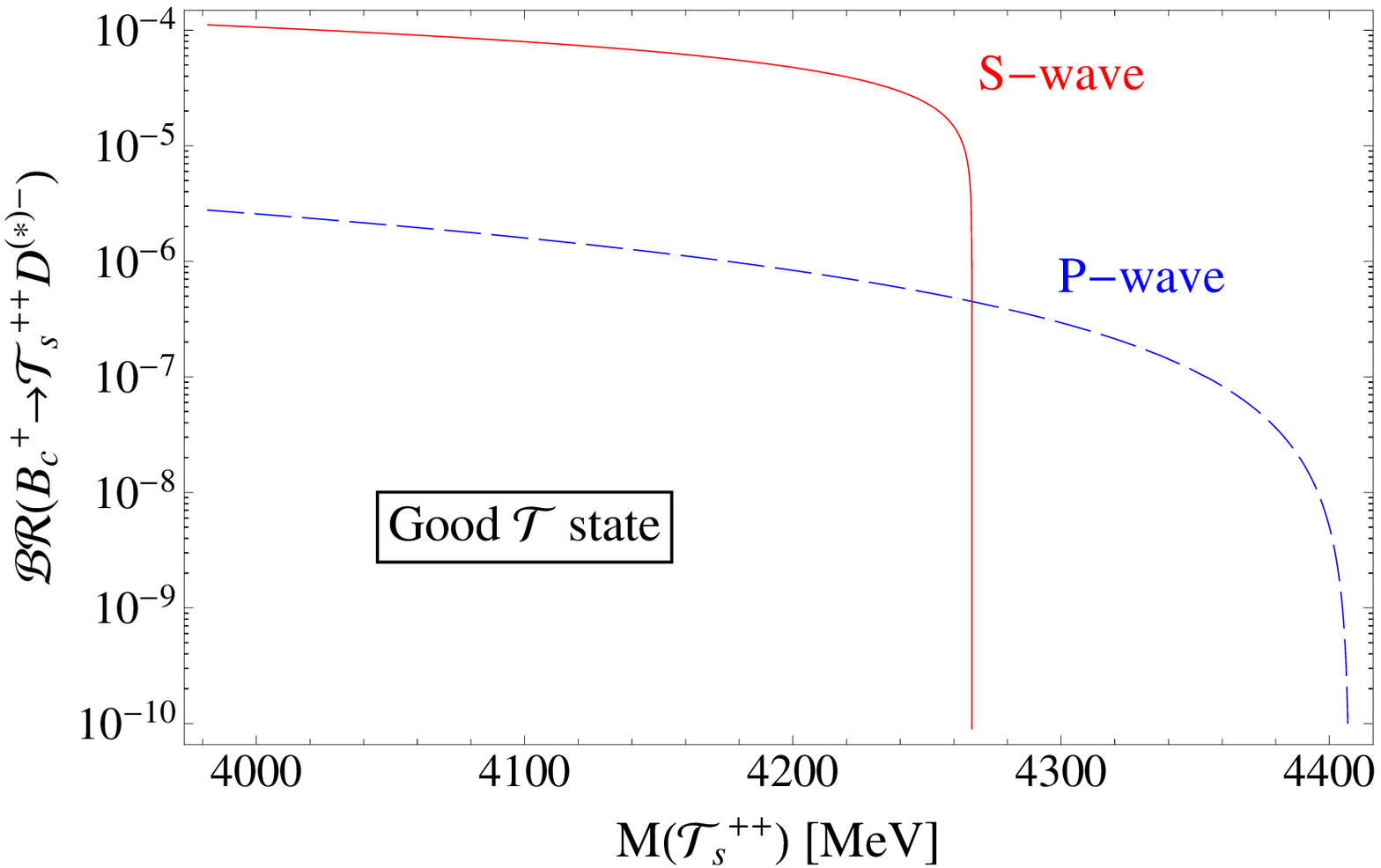}
\hspace{1em}
\includegraphics[scale=0.5]{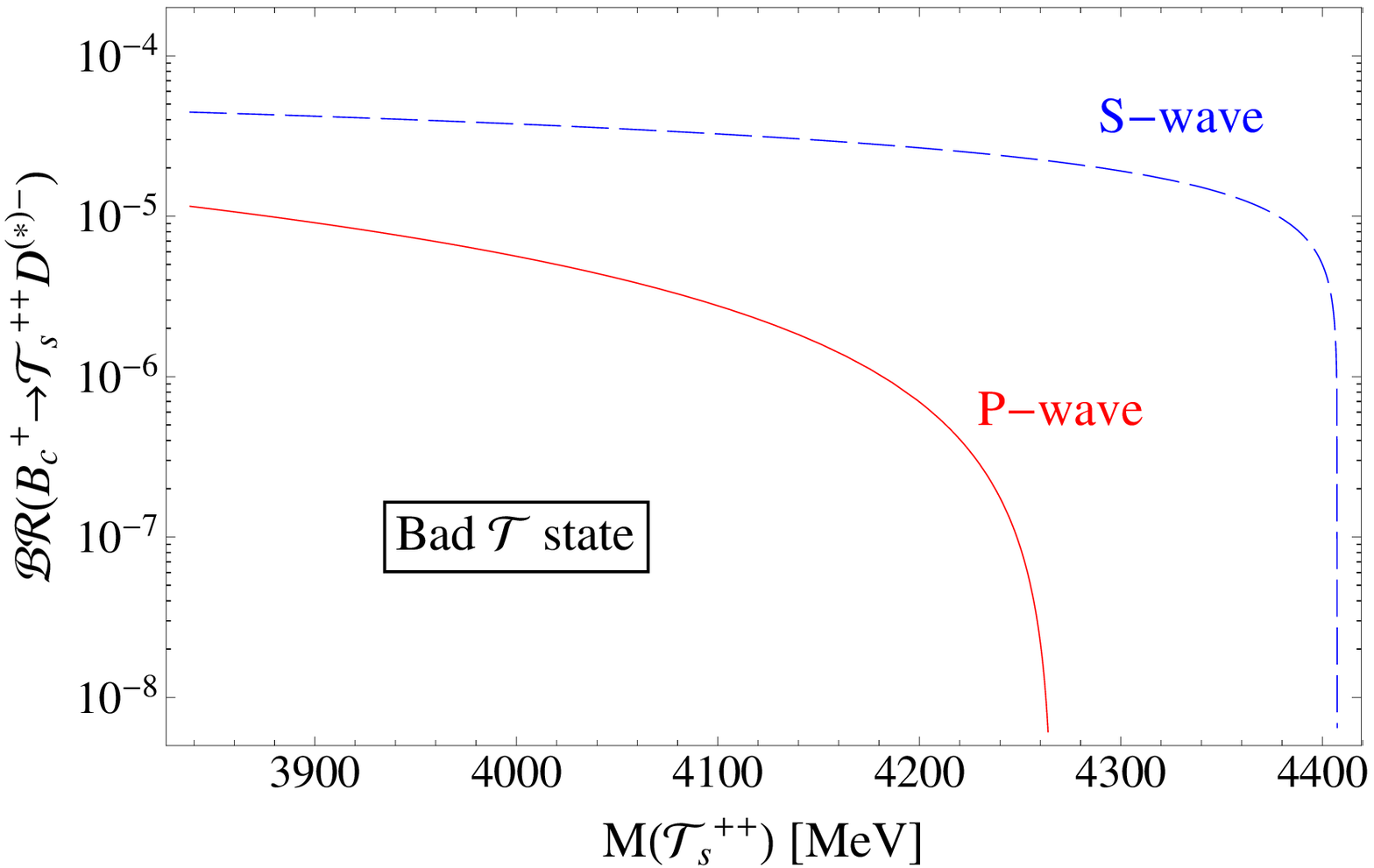}
\caption{Branching ratios for the production of $B_c^+\to\mathcal{T}_s^{++}\,D^{-}$ (dashed) and $B_c^+\to\mathcal{T}_s^{++}\,D^{*-}$ (solid) for the good $1^+$ state (left panel) and for the bad $0^+$ state (right panel) as a function of the mass of $\mathcal{T}_s^{++}$, in the above-threshold region. }
\label{fig:produzione_Ts}
\end{figure}
Therefore, from Eqns.~\eqref{eq:B0} and~\eqref{eq:matrix}, we can extract the effective couplings from the experimental values of the branching ratios
\begin{subequations}\label{eq:couplings}
\begin{align}
g_{B^0}&=\left(4\pm1\right)\times 10^{-3}\,\text{ MeV} \\
g_{B^+}&=\left(5\pm3\right)\times 10^{-3}\,\text{ MeV} 
\end{align}
\end{subequations}
which are compatible within errors.

As far as the $B_c^+\to \mathcal{T}_s^{++}D^{(*)-}$ process is concerned we can consider four different situations, \emph{i.e.} bad ($0^+$) and good ($1^+$) $\mathcal{T}$ states produced in association with either a $D^-$ or a $D^{*-}$. Depending on which of this four combinations one takes into account, the $B_c^+$ decay can take place in $S$- or $P$-wave. In particular the amplitudes can be written as
\begin{subequations}\label{eq:Tmatrix}
\begin{align}
\mathcal{A}_{B_c^+}^{(G)}\left[B_c^+ \to \mathcal{T}_s^{(G)++}\left(q,\epsilon\right)\,D^{*-}\left(p,\eta\right)\right]_\text{$S$-wave}&=g_{B_c^+}\left(\epsilon \cdot \eta\right) \\
\mathcal{A}_{B_c^+}^{(G)}\left[B_c^+ \to \mathcal{T}_s^{(G)++}\left(q,\epsilon\right)\,D^{-}\left(p\right)\right]_\text{$P$-wave}&=\frac{g_{B_c^+}}{M_{B_c^+}}\left(\epsilon \cdot p\right)\\
\mathcal{A}_{B_c^+}^{(B)}\left[B_c^+ \to \mathcal{T}_s^{(B)++}\left(q\right)\,D^{-}\left(p\right)\right]_\text{$S$-wave}&= g_{B_c^+} \\
\mathcal{A}_{B_c^+}^{(B)}\left[B_c^+ \to \mathcal{T}_s^{(B)++}\left(q\right)\,D^{*-}\left(p,\eta\right)\right]_\text{$P$-wave}&=\frac{g_{B_c^+}}{M_{B_c^+}}\left(\eta \cdot q \right)
\end{align}
\end{subequations}
where $g_{B_c^+}$ is an effective coupling with the dimension of a mass.
In particular we choose for $g_{B_c^+}$ the mean value of the couplings in Eq.~\eqref{eq:couplings}. Plugging the amplitudes~\eqref{eq:Tmatrix} into Eq.~\eqref{eq:width_tantalo} we can give an estimate of the branching ratio for the production of the $\mathcal{T}_s^{++}$ from the decays of the $B_c^+$. The results are shown in Fig.~\ref{fig:produzione_Ts}.

As one can see, in order to obtain an $S$-wave decay  (which is reasonably large) for the production of the $\mathcal{T}_s^{++}$ in the good configuration, one must require the production to be associated with a $D^{*-}$, while for the bad configuration it must be associated to a $D^-$. In particular, if the $\mathcal{T}$ is near the threshold, the branching ratios for such decays would only be one order of magnitude smaller than the recently observed decays $B_c^+ \to J/\psi\,D_s^{(*)+}$~\cite{lhcbjpsids}.

\begin{table}[tb!]
\centering
\begin{tabular}{c|cc}
\hline\hline
& Bottom quark decays & \\
\hline
Starting baryon & $b\to c\bar u d$  \hspace{1em}$(\mathcal{O}(\lambda^2))$ & $b\to c\bar u s$  \hspace{1em}$(\mathcal{O}(\lambda^3))$\\
\hline
& $p \mathcal{T}^0 \to p D^{0} D^{0}$ & $\Sigma^+ \mathcal{T}^0\to \Sigma^+ D^{0} D^{0}$ \\
$\Xi_{bc}^+\;[bcu]$ & $n \mathred{\mathcal{T}^+} \to n \mathred{D^{0} D^{+}}$ & $\Lambda^0(\Sigma^0) \mathred{\mathcal{T}^+} \to \Lambda^0(\Sigma^0) \mathred{D^{0} D^{+}}$ \\
& $\Lambda^0(\Sigma^0) \mathred{\mathcal{T}^+_s} \to\Lambda^0(\Sigma^0) \mathred{D^{0} D_s^{+}}$ & $\Xi^0 \mathred{\mathcal{T}_s^+} \to \Xi^0 \mathred{D_s^{+} D^{0}}$ \\
\hline
& $n \mathcal{T}^0 \to n D^{0} D^{0}$ &  $\Lambda^0(\Sigma^0) \mathcal{T}^0 \to \Lambda^0(\Sigma^0) D^{0} D^{0}$ \\
$\Xi_{bc}^0\;[bcd]$ & $\Delta^- \mathred{\mathcal{T}^+}\to \Delta^- \mathred{D^{+} D^{0}}$ & $\Sigma^- \mathred{\mathcal{T}^+} \to \Sigma^- \mathred{D^{+} D^{0}}$ \\
&  $\Sigma^- \mathred{\mathcal{T}_s^+} \to \Sigma^- \mathred{D_s^{+} D^{0}}$ & $ \Xi^- \mathred{\mathcal{T}_s^+} \to \Xi^- \mathred{D_s^{+} D^{0}}$ \\
\hline
& Same final states as $[bcd]$ & $\Xi^0 \mathcal{T}^0 \to \Xi^0 D^{0} D^{0}$\\
$\Xi_{bcs}^0\;[bcs]$ & with $b\to c\bar u s$ & $\Xi^- \mathred{\mathcal{T}^+} \to \Xi^- \mathred{D^{+} D^{0}}$ \\
& (they differ by just $d\leftrightarrow s$) & $\Omega^- \mathred{\mathcal{T}_s^+} \to \Omega^- \mathred{D_s^{+} D^{0}}$\\
\hline\hline
\end{tabular}
\caption{Possible final doubly charmed states produced by the proposed baryonic decays. With $\mathcal{O}(\lambda^n)$ we indicate the power of the Cabibbo suppression of the considered process. It is understood that each baryon and meson can also be found in its excited state, depending on the $J^P$ quantum numbers of the $\mathcal{T}$ particle considered -- see again Tab.~\ref{tab:decays}. We have marked in \red{red} good tetraquarks, and their decay products. In this case at least one charmed meson must be vectorial.} \label{table_baryon}
\end{table}

\subsection{\texorpdfstring{Production from $\Xi_{bc}$ baryons}{Production from Xi(bc) baryons}}

\begin{figure}[t]
\centering
\begin{picture}(160,100)(0,0)
\ArrowLine(10,95)(110,95)
\ArrowLine(10,45)(110,45)
\ArrowLine(10,55)(110,55)

\Photon(30,95)(70,75){1}{5}

\ArrowLine(70,75)(110,65)
\ArrowLine(110,85)(70,75)

\Gluon(30,45)(70,25){2}{7}

\ArrowLine(70,25)(110,15)
\ArrowLine(110,35)(70,25)

\Text(0,95)[cc]{$b$}
\Text(0,55)[cc]{$c$}
\Text(0,45)[cc]{$q$}

\Text(27,88)[cc]{$\lambda^2$}

\Text(120,95)[cc]{$c$}
\Text(120,84)[cc]{$\bar u$}
\Text(120,66)[cc]{$d$}
\Text(120,55)[cc]{$c$}
\Text(120,45)[cc]{$q$}
\Text(130,35)[cc]{$\bar u,\bar d, \bar s$}
\Text(130,15)[cc]{$u, d, s$}
\end{picture}
\begin{picture}(160,100)(0,0)
\ArrowLine(10,95)(110,95)
\ArrowLine(10,45)(110,45)
\ArrowLine(10,55)(110,55)

\Photon(30,95)(70,75){1}{5}

\ArrowLine(70,75)(110,65)
\ArrowLine(110,85)(70,75)

\Gluon(30,45)(70,25){2}{7}

\ArrowLine(70,25)(110,15)
\ArrowLine(110,35)(70,25)

\Text(0,95)[cc]{$b$}
\Text(0,55)[cc]{$c$}
\Text(0,45)[cc]{$q$}

\Text(27,88)[cc]{$\lambda^2$}
\Text(65,70)[cc]{$\lambda$}

\Text(120,95)[cc]{$c$}
\Text(120,84)[cc]{$\bar u$}
\Text(120,66)[cc]{$s$}
\Text(120,55)[cc]{$c$}
\Text(120,45)[cc]{$q$}
\Text(130,35)[cc]{$\bar u,\bar d, \bar s$}
\Text(130,15)[cc]{$u, d, s$}
\end{picture} 

\caption{Feynman diagrams for the production from $\Xi_{bc}$. For each baryon we can have $b\to c\bar u d$ (left) or $b\to c\bar u s$ (right). We consider here three different baryons, for $q=\{u,d,s\}$.}\label{diagrammi_baryon}
\end{figure}
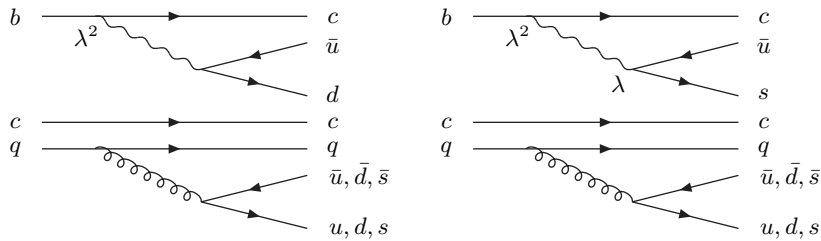

Heavy baryon production is expected to have high rates at LHC. In particular, LHC should be able to perform precise measurements of doubly heavy baryons, either with two charm quarks or one charm and one bottom quark. Our doubly charmed tetraquarks $\mathcal{T}$ could be observed in the decay of bottom-charm baryons, $\Xi_{bc}$.
Although these particles have not been discovered yet, we have taken into account this channel because of the characteristic signature of final states with baryons.
We consider here three different initial baryon species: $[bcu]\, (\Xi_{bc}^+)$, $[bcd]\,(\Xi_{bc}^0)$ and $[bcs]\,(\Xi_{bcs}^0)$. For each of them the bottom quark can decay into two different channels: $b\to c\bar u d$ and $b\to c\bar u s$, where the second channel is suppressed with respect to the first one by an additional Cabibbo angle factor. We have two possible kinds of Feynman diagrams, as reported in Fig.~\ref{diagrammi_baryon}. 

The list of the possible final states produced by these three baryons for each bottom quark decay are shown in Tab.~\ref{table_baryon}. In the table, every meson produced by the $\mathcal{T}$-decay can be found also in its excited state. However, when possible, it is preferable to choose the good (scalar) diquark configuration for the $[\bar q_1 \bar q_2]$ pair, as it leads to a lighter $\mathcal{T}$ state.

One must also notice that no doubly charged $\mathcal{T}$ can be produced by baryons if only one quark-antiquark pair creation is taken into account.

\section{Conclusions}
Doubly charmed $\mathcal{T}$ particles are a straightforward consequence of the constituent diquark-antidiquark model, and have good chances to be thoroughly studied on the lattice. If discovered, such particles would be almost full-proof tetraquarks. This is particularly true for the doubly charged ones. In this paper we have presented a first exploration of the above-threshold $\mathcal{T}$ particles phenomenology with the aim of stimulating their experimental search at the LHC. We have provided a list of decay modes and rates, together with production mechanisms, especially from $B_c$ mesons and $\Xi_{bc}$ baryons. Upcoming lattice studies will hopefully help to further constrain the identikit of the $\mathcal{T}$ sector of multiquarks.

\section*{Acknowledgements}
We wish to thank G. Carboni, R. Faccini, A. Palano and M. Sokoloff for stimulating discussions on the experimental aspects of $\mathcal{T}$ particles detection. We also thank F. Piccinini for useful comments.

\appendix

\section{Hadronic resonances on the lattice}
\label{sec:applatt}

Resonances are essentially a dynamical phenomenon and their study, even on the lattice, requires an understanding of the matrix elements of the scattering matrix, a difficult theoretical problem that has not been solved in general~\cite{Maiani:1990ca}. In Ref.~\cite{Luscher:1986pf} L\"uscher has been able to establish a connection between the two-particle discrete eigenvalues of the finite volume QCD Hamiltonian and the matrix elements of the \emph{infinite volume} scattering matrix under the inelastic threshold (see also Refs.~\cite{Rummukainen:1995vs,Lin:2001ek,Bedaque:2004kc,deDivitiis:2004rf,Kim:2005gf,Christ:2005gi,Bernard:2008ax,Bernecker:2011gh,Mohler:2012nh} for a largely incomplete list of references). Here we briefly review the results of Ref.~\cite{Luscher:1986pf} by using the pure tetraquark state $\mathcal{T}^{++}_s=[cc][\bar s\bar d]$ as an example. 

Should the $\mathcal{T}^{++}_s$ be a $0^+$ state, it could show up as a resonance in the $D^+ D_s^+$ channel. We shall assume that the mass of the resonance is sufficiently close to the $M_{D}+M_{D_s}$ threshold. More precisely, with exact isospin symmetry, we require 
\begin{equation}
M_{D}+M_{D_s} < 
M_{\mathcal{T}^{++}_s} <
M_{D}+M_{D_s}+2M_\pi 
\label{eq:conditionmass}
\end{equation}
The finite volume eigenvalues $W(k)$ corresponding to the two-particle states $\ket{D^+ D_s^+}$ at zero total momentum (center of mass frame) in the elastic scattering region, Eq.~\eqref{eq:conditionmass}, will be given by
\begin{equation}
W(k)=\sqrt{M_D^2+\vert \vec k \vert^2}+\sqrt{M_{D_s}^2+ \vert \vec k \vert^2}
\label{eq:twoscatt}
\end{equation}
In absence of interaction, on a finite volume and by imposing periodic boundary conditions on the quark fields, the spatial momentum $\vec k$ would assume the values $L \vec k=2\pi \vec n$ where $\vec n$ is a 3-vector with integer components. The energy of the interacting particles in the elastic scattering region is still given by Eq.~\eqref{eq:twoscatt} but the momentum $k=\vert \vec k \vert$ is the solution of a complicated non-linear equation involving the phase-shifts of the different angular momentum components. By assuming that only the $S$-wave phase-shift $\delta_S(k)$ is different from zero, the quantization condition can be written in the following compact form
\begin{equation}
n\pi-\delta_S(k)=\phi(k L/2 \pi)\, ,
\qquad n\in \mathbb{Z}
\label{eq:quanttwopt}
\end{equation}
where $\phi(x)$ is a known kinematical function given in Ref.~\cite{Luscher:1986pf}. 

The condition in Eq.~\eqref{eq:conditionmass} is needed because, 
if the energy of the two scattering particles is above the inelastic threshold ($D^+ D_s^+\to D^+D_s^+ +2\pi$), Eq.~\eqref{eq:quanttwopt} is not valid anymore and its generalization requires a substantial 
amount of theoretical work, though some steps in this direction have recently been performed~\cite{Hansen:2012tf,Briceno:2012rv}. 
On the other hand, the remarkable result of Eq.~\eqref{eq:quanttwopt} allows  to calculate the resonance mass if the condition of Eq.~\eqref{eq:conditionmass} is satisfied. On the lattice one can compute the energy $W(k)$ and obtain the corresponding phase shift $\delta_S(k)$  by using Eq.~\eqref{eq:quanttwopt}. 
The dependence of $\delta_S(k)$ with respect to the energy can be 
studied by changing the volume at fixed coupling and/or by changing 
the boundary conditions of the meson interpolating operators. If there 
is a resonance in this channel, at the value $k_\star$ such that 
$M_{\mathcal{T}^{++}_s}=W(k_\star)$ the scattering phase passes trough $\pi/2$.

We stress once again that the calculation strategy discussed above is possible, 
at least in principle, for pure tetraquark states because of the very peculiar 
assignment of the flavor quantum numbers of these hypothetical 
particles. For example, the same strategy would not apply to the 
calculation of the $J/\psi$ mass. The reason is that below 
$M_{J/\psi}$ there are plenty of open scattering channels, including
many particle states, and Eq.~\eqref{eq:quanttwopt} cannot be used anymore.
\begin{figure}[t]
\begin{center}
\includegraphics[width=0.25\textwidth]{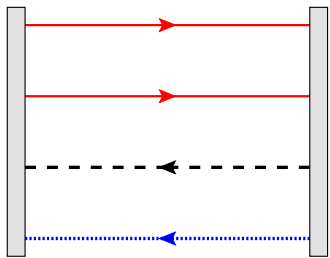}
\caption{Single diagram contributing to the propagation of a pure tetraquark state with
flavor quantum numbers \mbox{$Q_d=-1$}, \mbox{$Q_s=-1$,} \mbox{$Q_c=+2$.} The two grey walls representa the time slices at $0$ and $t$.  The red solid lines correspond to the two charm quark propagators, the black dashed line to the down quark propagator and the blue dotted line to the strange quark propagator. The fermionic lines are connected from all possible gluon exchanges and, consequently, include the contributions of sea quark loops.}\label{fig:purecontractions} 
\end{center}
\end{figure}
\begin{figure}[t]
\begin{center}
\includegraphics[width=0.25\textwidth]{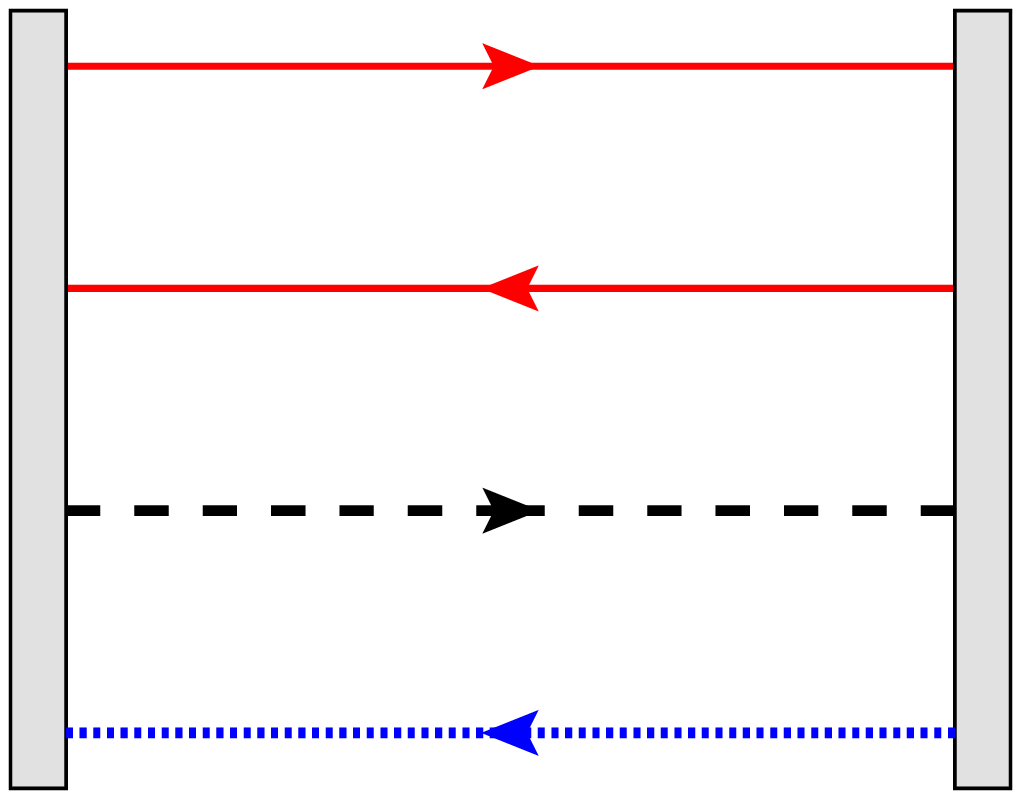}\hspace{1truecm}
\includegraphics[width=0.25\textwidth]{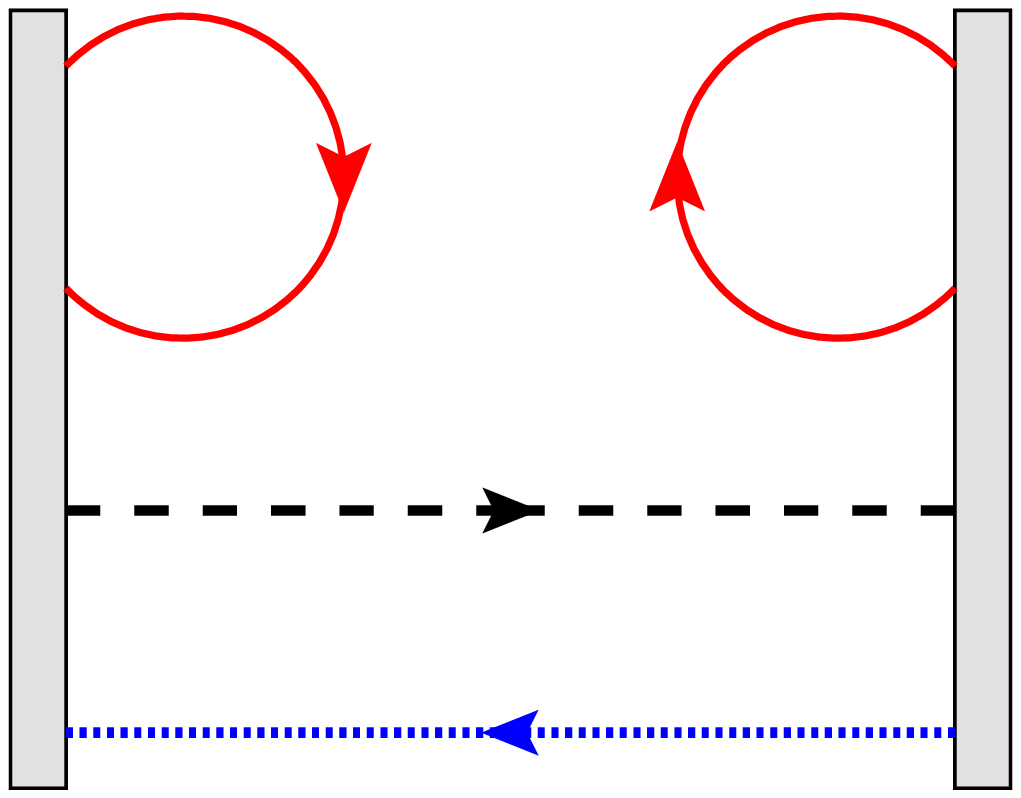}
\caption{The two diagrams contributing to the scattering of two mesons with flavor
quantum numbers $Q_d=1,\ Q_s=-1$, for example $D_s^+D^-$. The two grey walls representa the time slices at $0$ and $t$. The red solid lines correspond to the two charm quark propagators, the black dashed line to the down quark propagator and the blue dotted line to the strange quark propagator. The fermionic lines are connected from all possible gluon exchanges and, consequently, include the contributions of sea quark loops. }\label{fig:tetracontractions}
\end{center}
\end{figure}

By integrating out the quark fields from the functional integral, hadronic correlators can be expressed in terms of quark propagators. The flavor structure of a pure tetraquark is such that a single diagram contributes to the relevant correlation functions. This is shown in Fig.~\ref{fig:purecontractions}. The situation would be different for the scattering of a $D_s^+$ and a
$D^-$, where a tetraquark $[cd][\bar c\bar s]$ could be observed. In this case ``annihilation'' diagrams are possible, as shown in
Fig.~\ref{fig:tetracontractions}, and one can easily understand that the
lightest state propagating in this channel has $Q_d=1,\ Q_s=-1$, {\it i.e.} a meson 
with the same flavor quantum numbers of a neutral kaon.

The numerical calculation of annihilation diagrams is very challenging (and in fact they are usually neglectd by introducing an uncontrollable systematics) and the fact that they are forbidden in our case makes our calculation technically simpler and more reliable.

On the other hand, the presence of additional diagrams in the $D_s^+ D^-$ scattering amplitude with respect to the $D_s^+ D^+$ case means that the interaction in the two channels is different.  Annihilation diagrams are present in all channels in which standard hadronic resonances have been observed so far and this might be a sign that they could play a crucial role in generating an attractive interaction sufficiently strong to allow for the formation of bound states. In order to settle this question we think that a systematic investigation of the meson-meson interaction generated by the diagram in Fig.~\ref{fig:purecontractions} has to be performed.
By using the lattice techniques sketched in this section it should be possible to study the scattering phases at different energy scales (by varying the quark masses) and for different spin/angular momentum, parity and charge conjugation quantum numbers.
In the most favourable situation, pure tetraquark states may show up in this investigation. This will be the subject 
of future work. 

After the first version of this manuscript were submitted, some papers appeared related to the present work. In Ref.~\cite{gnocca} one can find a detailed investigation, based on the L\"uscher method, about the emergence of $X(3872)$ and $Z_c(3900)$ lines in the QCD spectrum.   
In Ref.~\cite{giapponese},  a preliminary work, on the kind of states we are directly interested in, has been presented with the HAL QCD method~\cite{hal}.  

\section{Calculation of color factors}\label{app:calcoli}
Let us first introduce a compact notation, \emph{i.e.} $\bar q_1^i q_{2i}\equiv \mathbb{1}_{\bar{q}_1 q_2}$ and $\bar{q}_1^i T^A_{ij}q_2^j\equiv T^A_{\bar{q}_1 q_2}$. With such a notation we have
\begin{align}
T^A_{{q}_1 q_2}\otimes T^A_{{q}_3 q_4}&=\frac{1}{2}\id_{{q}_1 q_4}\otimes\id_{{q}_3 q_2}-\frac{1}{6}\id_{{q}_1 q_2}\otimes\id_{{q}_3q_4}\label{eq:4}  \\
\id_{{q}_1q_2}\otimes\id_{{q}_2q_1}&=\tr(\id)\nonumber \\
T^A_{{q}_1q_2}\otimes T^B_{{q}_2q_1}&=\tr\left(T^AT^B\right)\nonumber \\
\id_{ q_1q_2}\otimes\id_{ q_2 q_3}&=\id_{ q_1 q_3} \nonumber
\end{align}

Let us start by considering the production of $\mathcal{T}$ as in Fig.~\ref{fig:produzione_barioni}. As the two charm quarks differ by the value of their momenta we distinguish them with the notation $c$ and $c^{\prime}$. The color structure of the final state of such a diagram is given by
\begin{equation}\label{eq:operatore}
\mathcal{O}\sim T^A_{\bar c c'}\otimes T^A_{\bar dd}\otimes \id_{\bar sc}=\frac{1}{2}\id_{\bar cd}\otimes\id_{\bar dc'}\otimes\id_{\bar sc}-\frac{1}{6}\id_{\bar cc'}\otimes\id_{\bar dd}\otimes\id_{\bar sc}\equiv \frac{1}{2}\mathcal{O}_1-\frac{1}{6}\mathcal{O}_2
\end{equation}
The final state of the reaction is given by
\begin{equation}
\ket{\mathcal{T}_s^{++}D^-}=\ket{\left(\epsilon^{ijk}c_ic'_j\epsilon^{lmk}\bar{s}_l\bar{d}_m\right)\otimes\id_{\bar cd}}=\ket{\id_{\bar sc}\otimes\id_{\bar dc'}\otimes\id_{\bar cd}}-\ket{\id_{\bar dc}\otimes\id_{\bar s c'}\otimes\id_{\bar cd}}
\end{equation}
The nomalization of the color wave function of such a state can be found using Eqns.~\eqref{eq:4}
\begin{align*}
\langle\mathcal{T}_s^{++}D^-|\mathcal{T}_s^{++}D^-\rangle =&
\langle\id_{\bar sc}\otimes\id_{\bar dc'}\otimes\id_{\bar cd}|\id_{\bar sc}\otimes\id_{\bar dc'}\otimes\id_{\bar cd}\rangle
-2\langle\id_{\bar sc}\otimes\id_{\bar dc'}\otimes\id_{\bar cd}|\id_{\bar dc}\otimes\id_{\bar s c'}\otimes\id_{\bar cd}\rangle \\
+&\langle\id_{\bar dc}\otimes\id_{\bar s c'}\otimes\id_{\bar cd}|\id_{\bar dc}\otimes\id_{\bar s c'}\otimes\id_{\bar cd}\rangle
=2 \tr\left(\id\right)^3-2 \tr\left(\id\right)^2=36
\end{align*}
Thus our normalized final state is given by

\begin{equation}\label{eq:stato}
\ket{\mathcal{T}_s^{++}D^-}=\frac{1}{6}\left(\ket{\id_{\bar sc}\otimes\id_{\bar dc'}\otimes\id_{\bar cd}}-\ket{\id_{\bar dc}\otimes\id_{\bar s c'}\otimes\id_{\bar cd}}\right)\equiv\frac{1}{6}\left(\ket{1}-\ket{2}\right)
\end{equation}

In order to evaluate the amplitude $\bra{B_c^+}\mathcal{O}\ket{\mathcal{T}_s^{++}D^-}$ we need to find the components of the operator $\mathcal{O}$ which have non-zero eigenvalues on the final state. Let us consider, for example, the action of the first term of Eq.~\eqref{eq:operatore} on the first term of Eq.~\eqref{eq:stato}
\begin{align}
\mathcal{O}_1\ket{1}&=\left(\id_{\bar cd}\otimes\id_{\bar dc'}\otimes\id_{\bar sc}\right)\ket{\id_{\bar sc}\otimes\id_{\bar dc'}\otimes\id_{\bar cd}}=\ket{\id_{\bar sc}\otimes\id_{\bar dc'}\otimes\id_{\bar cd}}
\end{align}
because the operator is exactly the right one for the considered state. For the second operator, $\mathcal{O}_2$, we have instead
\begin{align}
\mathcal{O}_2\ket{1}&=\left(\id_{\bar cc'}\otimes\id_{\bar dd}\otimes\id_{\bar sc}\right)\ket{\id_{\bar sc}\otimes\id_{\bar dc'}\otimes\id_{\bar cd}}\nonumber \\
&=\left[2\left(\id_{\bar sc}\otimes T^A_{\bar dc'}\otimes T^A_{\bar cd}\right)+\frac{1}{3}\left(\id_{\bar sc}\otimes\id_{\bar dc'}\otimes\id_{\bar cd}\right)\right]\ket{\id_{\bar sc}\otimes\id_{\bar dc'}\otimes\id_{\bar cd}} \nonumber \\
&=\frac{1}{3}\ket{\id_{\bar sc}\otimes\id_{\bar dc'}\otimes\id_{\bar cd}}
\end{align}
where the first term of the second equality is zero because the quarks are in the ``right'' combination but in the ``wrong'' color representation. Consequently we have $\mathcal{O}\ket{1}=\left(\frac{1}{2}-\frac{1}{18}\right)\ket{1}=\frac{4}{9}\ket{1}$. We now apply an analogous procedure for the second state of Eq.~\eqref{eq:stato} and obtain
\begin{equation}
\bra{B_c^+}\mathcal{O}\ket{\mathcal{T}_s^{++}D^-}=\frac{1}{6}\left[\frac{4}{9}\langle B_c^+\ket{ \id_{\bar sc}\otimes\id_{\bar dc'}\otimes\id_{\bar cd}}-\frac{4}{27}\langle B_c^+\ket{\id_{\bar dc}\otimes\id_{\bar s c'}\otimes\id_{\bar cd}}\right]
\end{equation}
We can notice that the two scalar products differ by just the exchange $c\leftrightarrow c'$ and we can therefore reasonably assume $\langle B_c^+\ket{ \id_{\bar sc}\otimes\id_{\bar dc'}\otimes\id_{\bar cd}}\simeq\langle B_c^+\ket{\id_{\bar dc}\otimes\id_{\bar s c'}\otimes\id_{\bar cd}}$. The complete decay width then reads
\begin{equation}
\Gamma(B_c^+\to\mathcal{T}_s^{++}D^{(*)-})=\frac{p^*(M_{B_c^+},M_{\mathcal{T}},M_{D^{(*)-}})}{8\pi M^2_{B_c^+}}{\left|\frac{1}{6}\cdot\frac{8}{27}\right|}^2{\left|\mathcal{A}_{B_c^+}\right|}^2
\end{equation}
where $\mathcal{A}_{B_c^+}=\langle B_c^+\ket{ \id_{\bar sc}\otimes\id_{\bar dc'}\otimes\id_{\bar cd}}$.

Analogously to the previous treatment, the color structure in the final state of $B^0 \to \bar \Lambda_c^-\,p$ is given by
\begin{equation}
\mathcal{O}\sim T^A_{\bar cd}\otimes T^A_{\bar uu'}\otimes \id_{\bar du}=\frac{1}{2}\id_{\bar cu'}\otimes\id_{\bar ud}\otimes\id_{\bar du}-\frac{1}{6}\id_{\bar c d}\otimes\id_{\bar u u'}\otimes\id_{\bar du}
\end{equation}

The normalized final state is, this time, more complex and is given by
\begin{align}
\ket{\bar{\Lambda}_c^- p}=\frac{1}{6}&\ket{\epsilon^{ijk}\bar u_i \bar d_j\bar c_k \epsilon^{\ell mn}u_\ell u_m'd_n} \nonumber \\
=\frac{1}{6}&\left[\ket{\id_{\bar uu}\otimes\id_{\bar du' }\otimes\id_{\bar cd}}-\ket{\id_{\bar uu}\otimes\id_{\bar dd}\otimes\id_{\bar cu'}}-\ket{\id_{\bar uu'}\otimes\id_{\bar du}\otimes\id_{\bar cd}}\right. \nonumber \\
&\left.+\ket{\id_{\bar uu'}\otimes\id_{\bar dd}\otimes\id_{\bar cu}}+\ket{\id_{\bar ud}\otimes\id_{\bar du}\otimes\id_{\bar cu'}}-\ket{\id_{\bar ud}\otimes\id_{\bar du'}\otimes\id_{\bar cu}}\right] \nonumber \\
\equiv\frac{1}{6}&\left[\ket{1}-\ket{2}-\ket{3}+\ket{4}+\ket{5}-\ket{6}\right]\label{eq:barioni}
\end{align}
where we used the well-known property for the product of two Levi-Civita tensors.
Again, we can apply $\mathcal{O}$ to the state in Eq.~\eqref{eq:barioni} searching for the components which have non-zero eigenvalues on the kets $\ket{i}$. We finally obtain
\begin{equation}
\mathcal{O}\ket{\bar{\Lambda}_c^-p}=\frac{1}{6}\left[-\frac{4}{27}\ket{2}+\frac{4}{9}\ket{5}-\frac{4}{27}\ket{6}\right]
\end{equation}

In close analogy to the case of the $\mathcal{T}$ we can reasonably set $\bra{B^0}i\rangle\simeq\bra{B^0}j\rangle\equiv\mathcal{A}_{B^0}$ for all $i,j$. Thus the decay width is given by
\begin{equation}
\Gamma(B^0\to\bar{\Lambda}_c^-p)=\frac{p^*(M_{B^0},M_{\Lambda_c^-},M_{p})}{8\pi M_{B^0}^2}{\left|\frac{1}{6}\cdot\frac{4}{27}\right|}^2{\left|\mathcal{A}_{B^0}\right|}^2
\end{equation}

\end{document}